\documentclass[a4paper]{article}

\usepackage{INTERSPEECH2022}
\usepackage{amsmath,graphicx, amsfonts}
\usepackage{multirow}
\usepackage{xcolor}
\usepackage{soul}
\usepackage{enumitem}
\usepackage[flushleft]{threeparttable}
\PassOptionsToPackage{hyphens}{url}\usepackage{hyperref}
\hypersetup{colorlinks=true}
\usepackage{changes}
\usepackage{todonotes}

\title{A Unified Cascaded Encoder ASR Model for Dynamic Model Sizes}
\name{Shaojin Ding$^{\star}$\thanks{$\star$ Equal Contribution. Listed in alphabetical order.}, Weiran Wang$^{\star}$, Ding Zhao$^{\star}$, Tara N. Sainath, Yanzhang He, Robert David, \\ Rami Botros, Xin Wang, Rina Panigrahy, Qiao Liang, Dongseong Hwang, Ian McGraw, \\ Rohit Prabhavalkar, Trevor Strohman}
\address{Google LLC, USA}
\email{\{shaojinding, weiranwang, dingzhao\}@google.com}

\begin{document}

\maketitle
\begin{abstract}

\noindent
In this paper, we propose a dynamic cascaded encoder Automatic Speech Recognition (ASR) model, which unifies models for different deployment scenarios. Moreover, the model can significantly reduce model size and power consumption without loss of quality. Namely, with the dynamic cascaded encoder model, we explore three techniques to maximally boost the performance of each model size: 1) Use separate decoders for each sub-model while sharing the encoders; 2) Use funnel-pooling to improve the encoder efficiency; 3) Balance the size of causal and non-causal encoders to improve quality and fit deployment constraints. Overall, the proposed large-medium model has 30\% smaller size and reduces power consumption by 33\%, compared to the baseline cascaded encoder model. The triple-size model that unifies the large, medium, and small models achieves 37\% total size reduction with minimal quality loss, while substantially reducing the engineering efforts of having separate models.

\end{abstract}

\section{Introduction \label{sec:intro}}
% \footnote{Preprint. Submitted to INTERSPEECH}  % TODO remove
End-to-end (E2E) models~\cite{wang2019overview, hannun2014deep, graves2012sequence, chorowski2015attention, dong2018speech} have gained popularity over the past few years, particularly for on-device automatic speech recognition (ASR), as they can achieve similar recognition performance compared to conventional hybrid systems~\cite{Golan16} at a fraction of the size. Over the past few years, developing an E2E model that surpasses conventional models in both quality and latency in diverse test conditions has been an active research area across many research groups~\cite{li2020comparison,Ryan19,CC18,KimHoriWatanabe17,JinyuLi2019,Zeyer2020}.

Recently, we presented an on-device E2E model based on a two-pass cascaded encoder which outperforms a conventional model in terms of word error rate (WER) on both search and long-tail queries, as well as endpointer latency metrics~\cite{sainath2021cascadedlm}. We further adapted the cascaded encoder to a small 1st-pass (50M parameters) large 2nd-pass (100M parameters) architecture to improve computational latency for both cloud and edge tensor processing units (TPUs), while maintaining quality~\cite{sainath2022improving}.

However, on-device ASR systems often require different model sizes for deployment to a variety of edge devices with different hardware constraints, e.g. mobile phones, home speakers, or cars. Even in the same device, different model sizes might still be required for various application constraints, e.g. a large model might be used for short-form applications (like voice search) to obtain the best quality, while a medium or a small model might be required for long-running applications (like dictation or video captioning) to maintain low power consumption. It is inefficient to train these different-sized models separately with duplicate efforts and high maintenance cost, especially for multiple languages.

To support such diversity of scenarios, we propose an approach by extending the cascaded encoder architecture in~\cite{sainath2021cascadedlm} to unify multiple size configurations in a single model during training. By only running a subset of the model layers at inference time, the model can be executed as different sizes with similar accuracies as the independently trained models of the corresponding sizes. This greatly reduces both the training overhead and the management complexity of deployment processes, and also allows run-time on-the-fly model size adjustment for variable resource usage.
Furthermore, we apply the following novel optimizations to improve quality, memory and latency: 1) Replace the shared decoder in the cascaded encoder model with separate decoders, which we will show is more robust to smaller encoder sizes; 2) Replace the stacking layer for downsampling in the causal encoder with a funnel-pooling layer to help reduce the size of the encoder \cite{dai2020funnel}; 3) Balance the size of causal and non-causal encoders to improve quality and fit deployment constraints.
We conduct extensive experiments on large scale tasks including voice search and dictation. Results show that our unified large-medium model achieves the same accuracy as the cascaded encoder baselines, with only about 70\% of model size, significantly reducing power consumption in the dictation task. Moreover, the unified large-medium-small model obtains minimal accuracy loss along with 37\% size reductions, compared to the upper-bounded individually trained models.

\textbf{Relation to prior work.} Several prior studies also explored the idea of jointly training ASR models with different sizes. The closest works to ours are~\cite{nagaraja2021collaborative,shi2021dynamic}, which investigated encoder and decoder weight sharing among large/medium/small models. However, all their encoder layers are non-causal, leading to significant latency increase at inference time. By contrast, our proposed model unifies both causal and non-causal layers, which makes it more efficient and flexible under different hardware constraints. More importantly, in these works, the model of each size has leveraged dedicated encoder layers that are not shared with other model sizes, which increases the overall model size.
However, as we have shown in the experiments, using smaller separate decoders avoids additional model size overhead and even allows the use of smaller encoders without any performance degradation. Secondly, \cite{nagaraja2021collaborative, shi2021dynamic, swaminathan2021codert} had additional distillation loss terms during the joint model training. 
In contrary, our preliminary experiments show that it is not straightforward to perform distillation between the causal layers and non-causal layers to improve the performance of causal layers, potentially due to the different right context; this direction is left as future work.
Lastly, compared with the alternative approach of model shrinking with sparsity networks~\cite{wu2021dynamic, yang2021omni}, our model is dense and requires no additional hardware support. Furthermore, it is more convenient to control the amount of right context in each size within our framework, and our training pipeline is much simpler, without the need for warm-starting a sparse model with a trained dense model.
\section{Method}
\vspace{-3pt}

In this section, we first introduce the proposed dynamic cascaded encoder model architecture, followed by the detailed descriptions of each of our novel designs. Finally, we present two specific dynamic cascaded encoder model architectures for practical applications.

\vspace{-3pt}
\subsection{Dynamic cascaded encoder model}
\vspace{-3pt}
\label{sec:dynamic_model}

\begin{figure}[t]
\centering
  \includegraphics[width=1\linewidth]{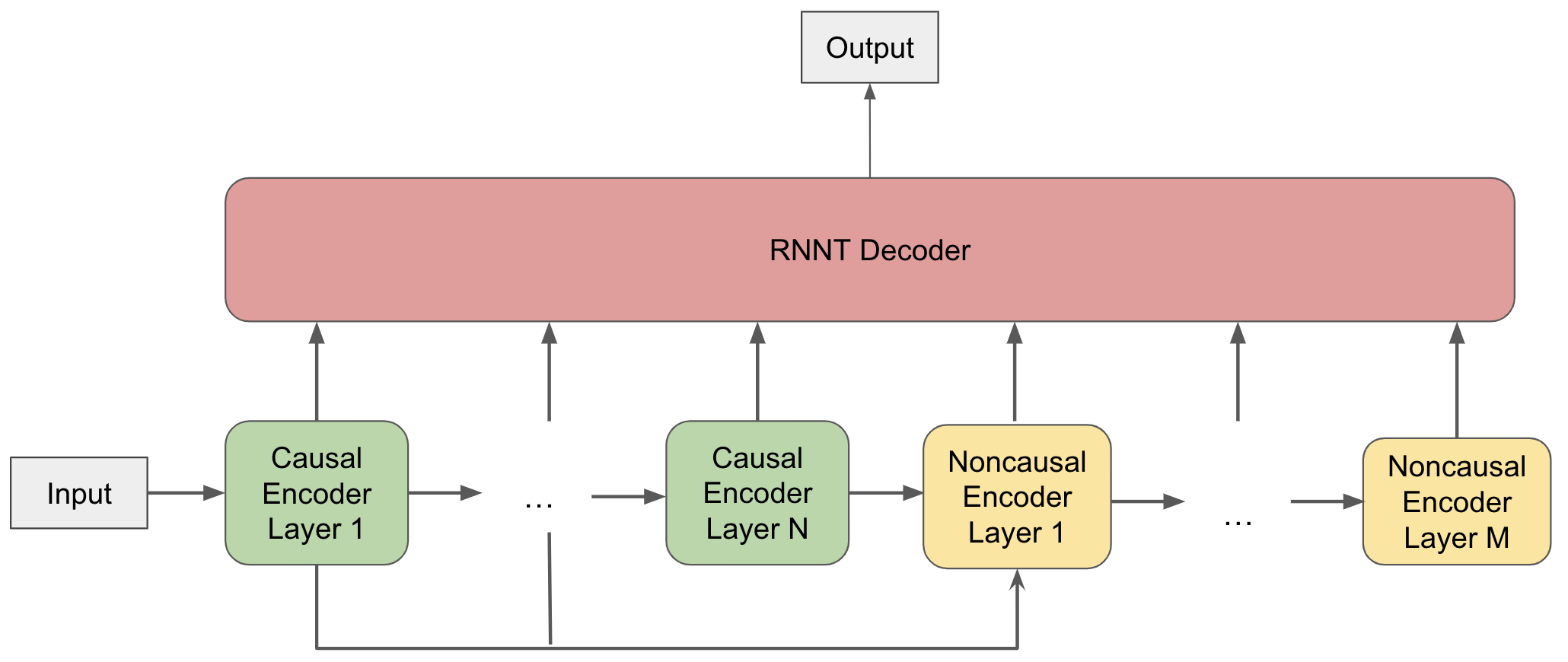}
  \vspace{-15pt}
  \caption{Dynamic cascaded encoder model structure.}
  \label{fig:model_structure}
  \vspace{-15pt}
\end{figure}

The baseline Conformer-based~\cite{gulati2020conformer} cascaded encoder model~\cite{sainath2022improving} is comprised of a causal conformer encoder with $N$ layers, followed by a non-causal conformer encoder~\cite{arun21cascade} with $M$ layers and an embedding RNN-T decoder~\cite{Rami21}. To improve the flexibility in unifying different models, we reformulate the cascaded model architecture to allow easy extractions of models with different sizes, as shown in Figure \ref{fig:model_structure}. 
In our model, each causal layer can be connected to the decoder or the first non-causal layer. We also allow connections from any non-causal layer to the decoder. From the super-net, we extract $K$ sub-models, each containing the first $n_k$ ($0 \leq n_k \leq N$) causal layers, and the first $m_k$ ($0 \leq m_k \leq M$) non-causal layers, which can be used under different model size and latency restrictions:

\vspace{-6pt}
\begin{equation}
    \mathbf{y}_k = \mathrm{Dec}(\mathrm{Enc}^{nc}_{k}(\mathrm{Enc}^{c}_{k}(\mathbf{x})))
\end{equation}

\noindent
where $\mathbf{x}$ and $\mathbf{y}_k$ denote the input and output of the $k$-th sub-model (all the sub-models have the same input). $\mathrm{Enc}^{c}_{k}$ is the causal encoder containing $n_k$ causal layers, $\mathrm{Enc}^{nc}_{k}$ is the non-causal encoder containing $m_k$ non-causal layers, and $\mathrm{Dec}$ is the shared decoder. Note that each of our sub-models does not have any dedicated encoder layer during training to minimize the total memory and storage cost in practice.

\vspace{-3pt}
\subsection{Separate decoders}
\vspace{-3pt}
\label{sec:separate-decoder}

The original cascaded encoder model~\cite{arun21cascade} uses a shared RNN-T decoder. The decoder works with a causal encoder in the first pass to provide streaming recognition results, and works with an additional non-causal encoder that sits on top of the causal encoder to provide more accurate final results, leveraging audio right context extracted by the noncausal encoder. Therefore, the same decoder has to deal with features of different context, and we observe tension between the performance of the passes as we try to reduce the model size, i.e., as we assign more loss weights for the causal pass to satisfy WER target, the accuracy of the non-causal pass degrades.

% \TNS{isnt this figure 2 now. you should reference it in the text.}
In this work, we propose to use smaller separate decoders in each sub-model, to better cope with the different context, and this significantly alleviates the tension between different sub-models:

\vspace{-6pt}
\begin{equation}
    \mathbf{y}_k = \mathrm{Dec}_k(\mathrm{Enc}^{nc}_{k}(\mathrm{Enc}^{c}_{k}(\mathbf{x})))
\end{equation}

\begin{figure}[t]
\centering
  \includegraphics[width=1\linewidth]{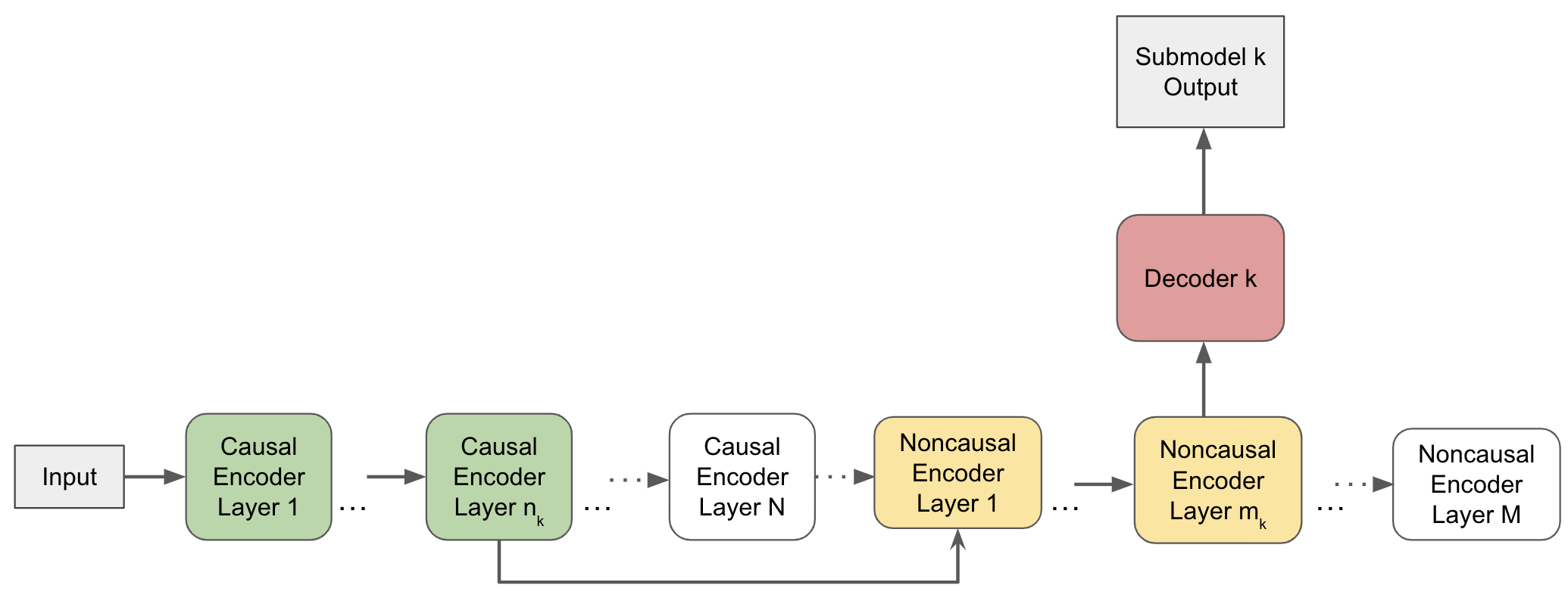}
  \vspace{-15pt}
  \caption{Sub-model with separate decoders in a Dynamic cascaded encoder model.}
  \label{fig:submodel}
  \vspace{-15pt}
\end{figure}

\noindent
Figure \ref{fig:submodel} shows an example of a sub-model with separate decoders: solid arrows are the connections used by this sub-model, and dotted arrows are connections used by other sub-models. 
As we will show in the experiments, empirically we can keep increasing the loss weight of the causal pass for better streaming results, without sacrificing performance of the non-causal pass. This allows us to use smaller separate decoders to replace the shared decoder, thus saving total memory cost and improving the inference speed of each sub-model.

\vspace{-3pt}
\subsection{Funnel-pooling layers}
\vspace{-3pt}

To reduce the overall computational cost, prior models usually use a stacking layer in the causal encoder to down-sample the input frame rate. The stacking layer concatenates features of two consecutive frames, and thus doubling the dimension of its output, which is used as input to the next attention layer and results in large amount of weight parameters in that layer.
However, it is extremely parameter-inefficient. To address the issue, we explore alternative down-sampling techniques. The most straight-forward substitution could be average pooling. However, using average pooling at the bottom layers usually introduce inevitable performance regressions~\cite{dai2020funnel}. Observing this, we propose to use funnel pooling~\cite{dai2020funnel} to down-sample the input frame rate, which has been shown to be able to preserve the model performance while reducing the frame rate in the middle of a sequential model.

Suppose we have a feature map $\mathbf{h}\in \mathbb{R}^{T\times D}$ as the input to a self-attention layer, where $T$ and $D$ denote the original sequence length and feature dimensions, respectively. We first create a down-sampled sequence of $\mathbf{h'}\in \mathbb{R}^{T'\times D}$ through average pooling:

\vspace{-10pt}
\begin{equation}
\label{eq:avg_pool}
  \mathbf{h'} = \mathrm{AvgPool}(\mathbf{h})
\end{equation}

\noindent
where $T' = T / 2$ in our case (down-sampled by a factor of 2). Instead of simply feeding $\mathbf{h'}$ to the self-attention, we only use $\mathbf{h'}$ as the query vector $\mathbf{q}$ in the self-attention layer. The key $\mathbf{k}$ and value vectors $\mathbf{v}$ are still based on the original input feature map $\mathbf{h}$:

\vspace{-10pt}
\begin{equation}
\label{eq:funnel_pool}
  \mathbf{h''} = \mathrm{SelfAttention}(\mathbf{q} = \mathbf{h'}, \mathbf{kv} = \mathbf{h})
\end{equation}

\noindent
where $\mathbf{h''}\in \mathbb{R}^{T'\times D}$ is the output feature maps.

\vspace{-3pt}
\subsection{Sub-model joint training}
\vspace{-3pt}

\label{sec:submodel-training}

We perform standard two-stage training as done in previous work. During maximum likelihood estimation training, we forward a minibatch through all sub-models and compute the loss of each sub-model:

\vspace{-6pt}
\begin{equation}
    L_k = \mathrm{Loss_{RNNT}}(\mathbf{y}_k)
\end{equation}

\noindent
and the losses for all sub-models are combined linearly,

\vspace{-6pt}
\begin{equation}
    L = \sum^{K}_{k=0}\lambda_k \cdot L_k
\end{equation}

\noindent
where $\lambda_k$ is the weight of the $k$-th sub-model, and all the weights sum to 1. After that, we continue fine-tuning the model with discriminative training using the MWER criteria~\cite{prabhavalkar2018}. For each step of MWER training, we randomly sample each sub-model with a probability equal to its loss weight, and use the sampled decoder to perform beam search on the minibatch to generate the top-4 hypotheses. The (full-sum) negative log-likelihood are computed for the hypotheses using the same sampled pass, and re-normalized in the top-4 space (so that the conditional "probabilities" sum to 1) to approximate the expected word error loss for minimization. 

\vspace{-3pt}
\subsection{Dynamic cascaded encoder model in practice}
\vspace{-3pt}

\begin{figure}[t]
\centering
  \includegraphics[width=1\linewidth]{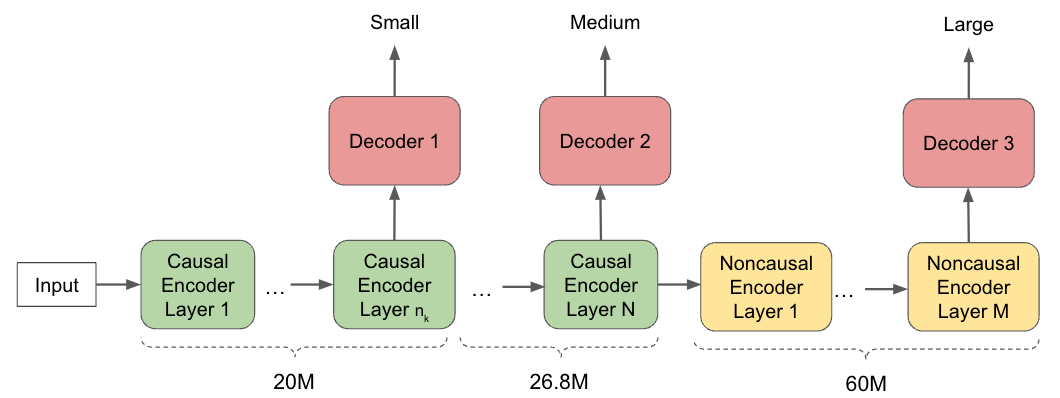}
  \vspace{-10pt}
  \caption{Triple-sized large-medium-small model.}
  \label{fig:three_sizes}
  \vspace{-15pt}
\end{figure}

With the flexibility of the dynamic cascaded encoder model, we establish a large-medium super-net and a large-medium-small super-net that work for most of the practical use cases. The large-medium super-net has a 46.8M causal encoder for the medium sub-model and an additional 60M non-causal encoder for the large pass, each having a 4.4M separate decoder. With the balanced size of causal and non-causal encoders, we show that it improves quality and fits deployment constraints better in Section~\ref{sec:exp_two_pass}. Our large-medium model only has around 70\% of model size, compared to the previous models in~\cite{sainath2021cascadedlm, sainath2022improving}.
Similarly, the large-medium-small super-net is comprised of a 20M causal encoder for the small sub-model, an additional 26.8M causal encoder for the medium sub-model, and a final 60M non-causal encoder for the large sub-model, as shown in Figure~\ref{fig:three_sizes}. The non-causal layer is only added to the large sub-model, because it requires fast hardware to catch up delays introduced by the right context, although it gives considerable quality gain. Each of the separate decoders also has 4.4M parameters. 

\section{Experimental setup}
\vspace{-3pt}

\subsection{Dataset}
\vspace{-3pt}

Similar to ~\cite{narayanan2019recognizing, sainath2020streaming}, all models are trained with $\sim$400k hours English audio-text pairs from multiple domains, such as YouTube and anonymized voice search traffic. YouTube data is transcribed in a semi-supervised fashion~\cite{liao2013large}. All other domains are anonymized and hand-transcribed. Our data handling abides by \textit{Google AI Principles}~\cite{googleaiprinciples}. We use a mixed-case word-piece vocabulary for all our experiments for on-device ASR to avoid a separate capitalization normalizer after decoding. This is different from previous studies~\cite{sainath2020streaming, sainath2021cascadedlm, sainath2022improving} that are conducted using lowercase wordepices for cloud-based E2E models. To avoid domain overfitting and increase data diversity, we apply two data augmentation techniques, including “multistyle training” (MTR)~\cite{kim2017mtr} and Spec-Augmentation~\cite{Park2019}.

During testing, we use the Voice Search (VS) test set and the Gboard Dictation Donation (Dictation) test set to evaluate the system performance. Voice Search contains around 12k voice search utterances, each having an average length of 5.5 seconds. Gboard Dictation Donation has 15k utterances and is collected as part of a voluntary program where users may choose to donate snippets of dictation speech to help improve speech models. Both search and dictation utterances are anonymized and hand-transcribed.

\vspace{-3pt}
\subsection{Implementation details}
\vspace{-3pt}

In our large-medium super-net, the causal encoder for the medium sub-model has seven 512-dimensional conformer layers (first three layers have no self-attention) with 23-frame left context per layer, and no right context to strictly prevent the model from using future inputs. The additional non-causal encoder for large pass has six 640-dimensional conformer layers, with additional 30-frame right context across six layers that processes 900ms speech from the future. All the self-attention layers have eight heads. Each separate RNN-T decoder is comprised of an 320-dimensional embedding prediction network and a 384-dimensional fully-connected joint network. We jointly train the super-net as described in Sec~\ref{sec:separate-decoder}, and we experimented with the weights in Section~\ref{sec:exp_sep_dec}. The large-medium-small super-net, has six 256-dimensional conformer layers for small sub-model, an additional six 512-dimensional causal conformer layers for the medium sub-model, and another six 640-dimensional non-causal layers for the large sub-model. The loss weights during joint model training are set to $[0.80, 0.15, 0.05]$ for small, medium, and large sub-models, respectively.

We use the 128-dimensional log Mel-filterbank enegies (extracted from 32ms window and 10ms shift) as the frontend feature, and then we stack the contiguous 4 frames, sub-sampled by a factor of 3, and append a 16-dimensional one-hot domain-ID vector~\cite{sainath2020streaming}. All our evaluations are running on an on-device inference pipeline, where we first convert the TensorFlow graphs to TensorFlow Lite format, and leverage the 8-bit post training quantization to further reduce the model file size. Additionally, we did not use any language model in our experiments, as this is orthogonal to the end-to-end model improvements.  The dictation power consumption is measured for recognizing a 14-minute continuous speech recording on a Pixel 6 mobile phone with the edge TPU on the Google Tensor chip.

\vspace{-3pt}
\section{Results}
\vspace{-3pt}

\begin{table*}[t]
\begin{center}
\caption{Comparing proposed large-medium models to cascaded encoder baselines. The numbers in grey indicate that the model cannot be used in practical application due to their high power consumption.}
\vspace{-8pt}
\label{table:2pass}
\resizebox{0.85\textwidth}{!}{
\begin{tabular}{|c|c|cc|cc|cc|cc|}
\hline
\multirow{2}{*}{Exp} & \multirow{2}{*}{Model} & \multicolumn{2}{c|}{VS WER} & \multicolumn{2}{c|}{Dictation WER} & \multicolumn{2}{c|}{Dictation Power (mW)} & \multicolumn{2}{c|}{Size (MB)} \\
 \cline{3-10} & & medium & large & medium & large & medium & large & medium & large  \\
 \hline
 B0 & Conf. cascaded encoder~\cite{sainath2021cascadedlm} & 6.9 & 5.8 & 5.8 & \textcolor[rgb]{.5,.5,.5}{5.3} & 272 & 410 & 120 & 152 \\
 B1 & Small 1st/Large 2nd~\cite{sainath2022improving} & 8.6 & 5.9 & 7.0 & \textcolor[rgb]{.5,.5,.5}{5.3} & 259 & 418 & 56 & 155 \\
 \hline
 E6 & Proposed large-medium model & 7.9 & 5.8 & 6.6 & 5.6 & 190 & 273 & 44 & 108 \\
\hline
\end{tabular}}
\vspace{-15pt}
\end{center}
\end{table*}

\begin{table*}[t]
\begin{center}
\caption{Comparing large-medium-small model with separately trained large/medium/small models.}
\vspace{-8pt}
\label{table:3pass}
\resizebox{0.65\textwidth}{!}{
\begin{tabular}{|c|c|ccc|c|}
\hline
\multirow{2}{*}{Exp} & \multirow{2}{*}{Model} & \multicolumn{3}{c|}{VS WER} & \multirow{2}{*}{Size (MB)} \\
 \cline{3-5} & & Small & Medium & Large &  \\
 \hline
 B2 & Separately trained models & 10.0 & 7.3 & 5.7 & 180 \\
 
 \hline
 E6 & Proposed large-medium-small model & 10.6 & 7.7 & 6.1 & 115 \\

\hline
\end{tabular}}
\vspace{-20pt}
\end{center}
\end{table*}

We conduct four sets of experiments to evaluate our proposed approach. First, we conduct two ablation studies verifying the impact of separate decoders and funnel pooling in the proposed dynamic cascaded encoder model, based on our large-medium model. Following this, we compare our best-performing large-medium model and large-medium-small model to the corresponding baseline methods, respectively, to show the effectiveness of our proposed approach.

\vspace{-3pt}
\subsection{Impact of separate decoders}
\vspace{-3pt}

\label{sec:exp_sep_dec}

We first examine the impact of the newly proposed separate decoders, by comparing with the previously used shared decoder approach~\cite{sainath2022improving}.
We provide the WERs on the VS testset in Table~\ref{table:sepdec1}. MWER training tends to reduce the WERs by similar amounts for both type of models, as shown in \textit{E4}.

\begin{table}[h!]
\begin{center}
\vspace{-4pt}
\caption{Evaluations on the effectiveness of separate decoders. We provide VS WERs (\%) for all models (E0 to E3 without MWER; E4 with MWER).}
\vspace{-7pt}
\label{table:sepdec1}
\resizebox{\columnwidth}{!}{
\begin{tabular}{|c|c|cc|cc|}
\hline
\multirow{2}{*}{Exp} & \multirow{2}{*}{medium/large weights} & \multicolumn{2}{c|}{Shared dec.} & \multicolumn{2}{c|}{Separate decs.} \\
 \cline{3-6} & &  medium & large & medium & large \\
  \hline
E0 & 0.6/0.4 & 9.0 & 6.5 & 9.0 &  6.1 \\
%  0.7/0.3 & 8.3 & 6.1 & 8.7 &  5.9 \\
E1 &  0.8/0.2 & 8.7 & 6.5 & 8.7 & 6.2 \\
E2 &  0.9/0.1 & 8.4 & 6.6 & 8.5 &  6.2 \\
E3 &  0.95/0.05 & 8.2 & 6.9 & 8.5 &  6.2 \\
\hline
E4 & 0.9/0.1 w/ MWER & 7.8 & 6.2 & 7.9 & 5.8 \\
\hline
\end{tabular}}
\vspace{-25pt}
\end{center}
\end{table}

As we skew the loss weight towards the small sub-model, shared decoder models do get improved accuracy for the small sub-model, and the WER reduces from 9.0\% to 8.2\% when the its weight increase from 0.6 to 0.95. However, this comes at the cost of a worse second pass, whose WER increase from 6.5\% to 6.9\%. 
In comparison, for models with separate decoders, as the medium sub-model WER decrease from 9.0\% to 8.5\%, the large sub-model WER only degraded by 0.1\% from 6.1\% to 6.2\%.
Therefore, we stick to the separate decoders setup with 0.9 vs 0.1 loss weights.

\vspace{-3pt}
\subsection{Impact of funnel pooling}
\vspace{-3pt}

To evaluate the effectiveness of funnel pooling, we compare it against two variants, i.e., using stacking and using average pooling for down-sampling. Results are shown in Table~\ref{table:funnel}. As we expect, the model with funnel pooling can achieve the same WERs as the model based on stacking. Additionally, comparing funnel pooling and average pooling, we do see a 0.2 WER regression in the model based on average pooling for both medium and large sub-models, further demonstrating the necessity of funnel pooling.

\begin{table}[h!]
\begin{center}
\vspace{-3pt}
\caption{Evaluations on the effectiveness of funnel pooling.}
\vspace{-7pt}
\label{table:funnel}
\resizebox{0.9\columnwidth}{!}{
\begin{tabular}{|c|c|cc|c|}
\hline
\multirow{2}{*}{Exp} & \multirow{2}{*}{Model} & \multicolumn{2}{c|}{VS WER} & \multirow{2}{*}{Size (MB)} \\
\cline{3-4} & & medium & large & \\

\hline
E4 & Stacking & 7.9 & 5.8 & 115 \\
\hline
E5 & Average pooling & 8.1 & 6.0 & 108\\

E6 & Funnel pooling & 7.9 & 5.8 & 108 \\
 \hline

\hline
%  \cline{3-6} & VS &  \\
\end{tabular}}
\vspace{-20pt}
\end{center}
\end{table}

\vspace{-3pt}
\subsection{Comparisons between the large-medium model and baseline cascaded encoder models}
\vspace{-3pt}

\label{sec:exp_two_pass}

After validating the use of separate decoders and funnel pooling, we discuss the performance of the large-medium model. We consider two conformer cascaded encoder baselines: (\textit{B0}) The original conformer cascaded encoder model in~\cite{sainath2021cascadedlm}, and (\textit{B1}) the small 1st/large 2nd conformer cascaded encoder model \cite{sainath2022improving} that is optimized for cloud TPU. 

Results are shown in Table~\ref{table:2pass}. Comparing between the two baselines, we confirm the medium sub-model degradation issue of model \textit{B1} (6.9 vs. 8.6), which is one of the motivations of this study. Our proposed model (\textit{E6}) can significantly mitigate the degradation and improve the first pass WER from 8.6 to 7.9. More importantly, \textit{E6} has a much smaller total model size (108MB) compared to the baselines ($\sim$30\% relative reduction), while retaining the large sub-model VS WER.  Besides quality-wise improvements, the proposed model also benefits in terms of the power consumption. When using \textit{B0} or \textit{B1} in recognizing continuous speech, although large sub-model has a better WER, we still rely on only the medium sub-model, since running the large sub-model leads to much higher power consumption (e.g., B0: 270mW vs. 410mW). By contrast, with the reduced model size, the large sub-model of \textit{E6} achieves similar power consumption to that of the baselines so that it can be used for long-running applications, while obtaining 0.2 and 1.4 absolute dictation WER reduction compared to the medium sub-models of \textit{B0} and \textit{B1} respectively.

\vspace{-3pt}
\subsection{Comparisons between the large-medium-small model and the separately trained models}
\vspace{-3pt}

Finally, we illustrate the capability of our triple-size model that unifies the large, medium, and small model production models. We compare it against a baseline (\textit{B2}) of separately trained large, medium, and small models.
\textit{B2} can be treated as an upper-bound to the proposed model, as there is no weight sharing and each size has a dedicated optimized model. 
Table~\ref{table:3pass} shows the results of the two models.
Compared to separately trained models, our unified model reduces 37\% model size with only a minimal WER regression, and the 6.1 WER on the large sub-model has already surpassed the quality of the server conventional model~\cite{sainath2020streaming}. The unified model allows us to use smaller sub-models to reduce model loading or computational latency during model cold-start or bursty audio situations, while switching to larger sub-models afterwards for better quality without increasing much memory, similar to~\cite{shi2021dynamic}. Also, it reduces the engineering efforts in model tuning and runtime optimizations, which is beneficial to large scale productionizations. 

\section{Conclusions}
\vspace{-3pt}

We have proposed a dynamic cascaded encoder ASR model based on separate decoders, which generalizes well to different model sizes, unifying the large, medium, and small models for different deployment scenarios. Moreover, the model can significantly reduce model size and power consumption compared to prior methods. Our experimental results confirmed that the separate decoders obtained a more promising performance compared to the shared decoder. In addition, with separate decoders, we showed that the efficiency of the encoders can be further improved via funnel pooling and deliberately designing between causal/non-causal encoder sizes, resulting in a 30\% smaller model size without any performance loss. Compared to baseline models, the proposed model reduces dictation power consumption on large sub-model by 33\%, which makes it possible to run inference with large sub-model for dictation with improved quality. Compared to separately trained large, medium, and small models, the proposed architecture achieves 37\% total size reduction, with slight performance degradations.

% \section{Acknowledgements}

\bibliographystyle{IEEEtran}

\bibliography{mybib}

\end{document}